\documentstyle[12pt,aaspp4]{article}

\begin{document}

\title{A Group of Red, Ly$\alpha$ Emitting, High Redshift 
Galaxies\footnotemark}

\author{Paul J. Francis}
\affil{University of Melbourne, School of Physics, Parkville, Victoria 3052,
Australia \\ e-mail pfrancis@physics.unimelb.edu.au}

\author{Bruce E. Woodgate}
\affil{NASA Goddard Space-Flight Center, Code 681, Greenbelt, MD 20771 \\
e-mail woodgate@uit.dnet.nasa.gov}

\and

\author{Anthony C. Danks}
\affil{Hughes STX, Goddard Space-flight Center, Code 683.0, Greenbelt,
MD 20771 \\ e-mail danks@iue.gsfc.nasa.gov}

\footnotetext{Based on observations carried out at the Anglo Australian
Telescope, Cerro Tololo Interamerican Observatory, and Siding Spring
Observatory}

\begin{abstract}

We have discovered two new high redshift ($z=2.38$)
galaxies, near the previously known $z=2.38$ 
galaxy 2139$-$4434 B1 (Francis et al.\markcite{F96} 
1996). 

All three galaxies are strong Ly$\alpha$ emitters, and have
much redder continuum colors ($I-K \sim 5$) than
other optically-selected high redshift galaxies. We hypothesize that 
these three galaxies are QSO IIs; radio-quiet counterparts of high 
redshift radio galaxies, containing concealed QSO nuclei. The red colors
are most easily modelled by an old ($> 0.5$ Gyr), 
massive ($> 10^{11} M_{\sun}$) stellar population. If true, this 
implies that at least one galaxy cluster of mass $\gg 3 \times 10^{11} 
M_{\sun}$ had collapsed before redshift five.

\end{abstract}


\section{Introduction}

The first substantial samples of radio-quiet, high redshift ($z>2$)
galaxies are now being assembled (eg. Steidel et al. 1996\markcite{s96}, 
Lanzetta et al. 1996\markcite{l96}, Lowenthal et al. 1997\markcite{lo97}). 
These galaxies have blue continuum colors (longward of the Ly$\alpha$ 
forest absorption), probably due to ongoing star formation.
In contrast, many high redshift radio galaxies have very red continuum
colors (eg. McCarthy 1993\markcite{mc93}, Dunlop et al. 1996\markcite{du96}).

Recently, a small number of radio-quiet high redshift galaxies have
been found which share the red colors of high redshift radio galaxies.
Francis et al. (1996, hereafter F96)\markcite{F96} identified a 
very red radio-quiet
Ly$\alpha$-emitting galaxy (2139$-$4434 B1) at redshift 2.38, associated
with a cluster of QSO absorption-line systems (Francis \& Hewett 1993, 
hereafter\markcite{FH93} FH93).  Hu \& Ridgway 
1994\markcite{hr94} also identified a small number of extremely red
sources, at least one of which is a radio-quiet high redshift galaxy 
(Graham \& Dey 1996\markcite{gd96}). If additional
red, radio-quiet high-redshift galaxies were found, it would imply that the 
distinctively red colors of high redshift radio galaxies are not caused
by the radio jets, and hence might be common in the early universe.

In this letter, we present the results of a search for more 
Ly$\alpha$-emitting galaxies in the field around 2139$-$4434 B1. We assume 
$H_0 = 100 h_{100} {\rm km\ s}^{-1}{\rm Mpc}^{-1}$, $h_{100} = 0.75$ 
(except where stated) and $q_0 = 0.5$.

\section{Observations}

The field around 2139$-$4434 B1 was searched for Ly$\alpha$-emitting 
galaxies by imaging through a narrow-band filter tuned to a
central wavelength of 411.6 nm,
the expected wavelength of Ly-$\alpha$ emission at B1's redshift 
(z=2.38). Images were obtained on the night of 1995 October 27th, using
a 2k$\times$2k backside thinned CCD of the original HST/STIS (Space 
Telescope Imaging
Spectrograph) design built by SITe, and installed in a Photometrics
designed camera mounted at the prime focus of the Anglo-Australian
Telescope (AAT). The doublet image corrector gave a field of view $11.7\arcmin$
on a side, which was centered midway between the two background QSOs
described in FH93. \markcite{FH93} Offband images were obtained through a 
$B$ filter; exposure times were 5400 seconds in the narrow-band and
3600 seconds in the broad-band. A narrow-band surface brightness limit of
$5.0 \times 10^{-20} {\rm \ W\ m}^{-2}{\rm arcsec}^{-2}$
($5.0 \times 10^{-17} {\rm erg\ cm}^{-2}{\rm s}^{-1}{\rm arcsec}^{-2}$)
was achieved ($5 \sigma$), corresponding to a flux limit of $9 \times
10^{-20}{\rm \ W\ m}^{-2}$ ($9 \times
10^{-17}{\rm erg\ cm}^{-2}{\rm s}^{-1}$) for unresolved sources (seeing was
$1.4\arcsec$); the filter passband width was 6.4 nm, corresponding to
a redshift range of $\Delta z = 0.05$. The $B$-band image reached a 
($5 \sigma$) limiting magnitude of 25.8 for unresolved sources.
After removal of cosmic rays and bad pixels, the effective
area surveyed was $70$ square arcmin.

Four candidate Ly$\alpha$ emitting galaxies were detected, with excess 
narrow-band fluxes requiring an emission-line of rest-frame equivalent width 
$> 10$ nm ($3 \sigma$). These included F96's\markcite{F96} narrow-band
excess sources B1 and B2, but their marginal candidate B3 was not confirmed.

Confirmation spectroscopy was obtained with the Low Dispersion Survey
Spectrograph (LDSS) on the AAT, on the nights of August 13th and 14th 1996.
A slit mask was designed to include all four candidates, with spare slits
placed over lower priority candidates. A total of 47,700 seconds
exposure was obtained, in $1.2\arcsec$ seeing, the the slits aligned with a 
position angle of 340$^{\circ}$. The wavelength range covered
spanned Ly$\alpha$ to \ion{He}{2} at rest-frame z=2.38, with a resolution of
$700 {\rm km\ s}^{-1}$.

\begin{figure}

\epsscale{0.7}
\plotone{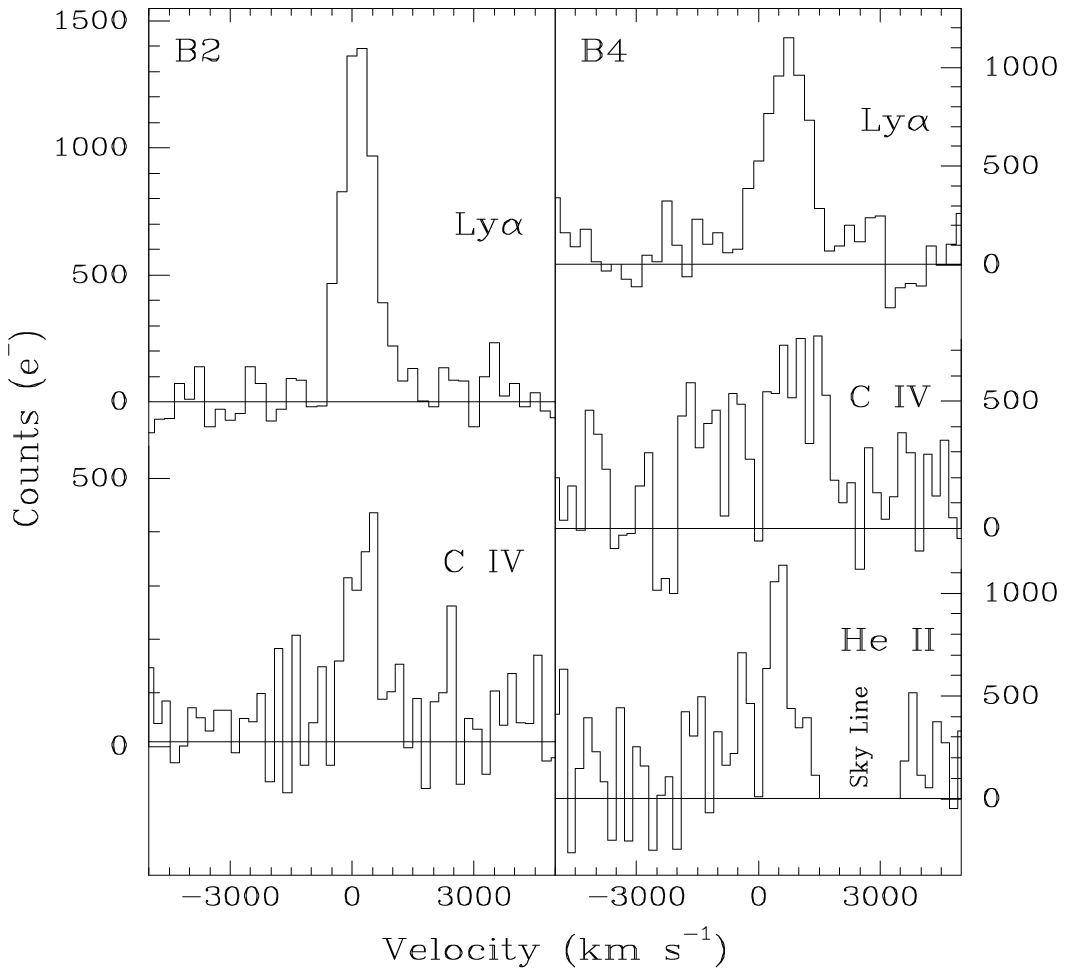}

\caption{Spectra of B2 and B4. Velocities are relative to
the centroid of the QSO absorption system in QSO 2139$-$4434, which is at
410.9 nm. Note that the red wing of \ion{He}{2} is clearly separated from the
sky line marked in the two-dimensional spectrum. The spectra have
been sky subtracted but not rebinned or flux calibrated. \label{Fig1}}

\end{figure}

A deep $R$-band image of the field was obtained with the prime focus CCD
at the CTIO 4-m telescope on October 19th 1995; exposure time was 4200
sec, and a ($5 \sigma$) limiting magnitude of 25.5 was reached. Deep
$I$-band imaging was obtained with LDSS on the AAT on November 11th 1996,
reaching magnitude 24.4 in $1\arcsec$ seeing. Optical photometric 
calibration was obtained on 1995 August 16th with the Siding Spring 1-m 
telescope. The $K_n$ imaging of F96\markcite{F96} was extended with the 
CASPIR camera on the Siding Spring 2.3-m
telescope, on August 12th and 13th 1995.
The positions of all candidates were checked against
the 13 cm and 20 cm radio maps of F96\markcite{F96}; none of the candidates 
are detected at a flux level of $0.27$ mJy at either frequency. $J$, $H$ and 
H$\alpha$ measurements were made from the data in F96\markcite{F96}.

\section{Results}

In addition to the previously confirmed B1, we detected \ion{C}{4} 
(154.9 nm) emission in 
F96's\markcite{F96} second candidate galaxy B2, thus confirming its redshift. 
We did not detect their marginally significant candidate B3, but we found 
and confirmed
a new Ly$\alpha$ source B4, for which we detect Ly$\alpha$, \ion{C}{4}, 
\ion{He}{2} (164.0 nm), and
(with $2 \sigma$ confidence) H$\alpha$. The spectra are shown in Fig~1. 
None of the other sources were
high redshift galaxies.

\begin{figure}

\epsscale{0.8}
\plotone{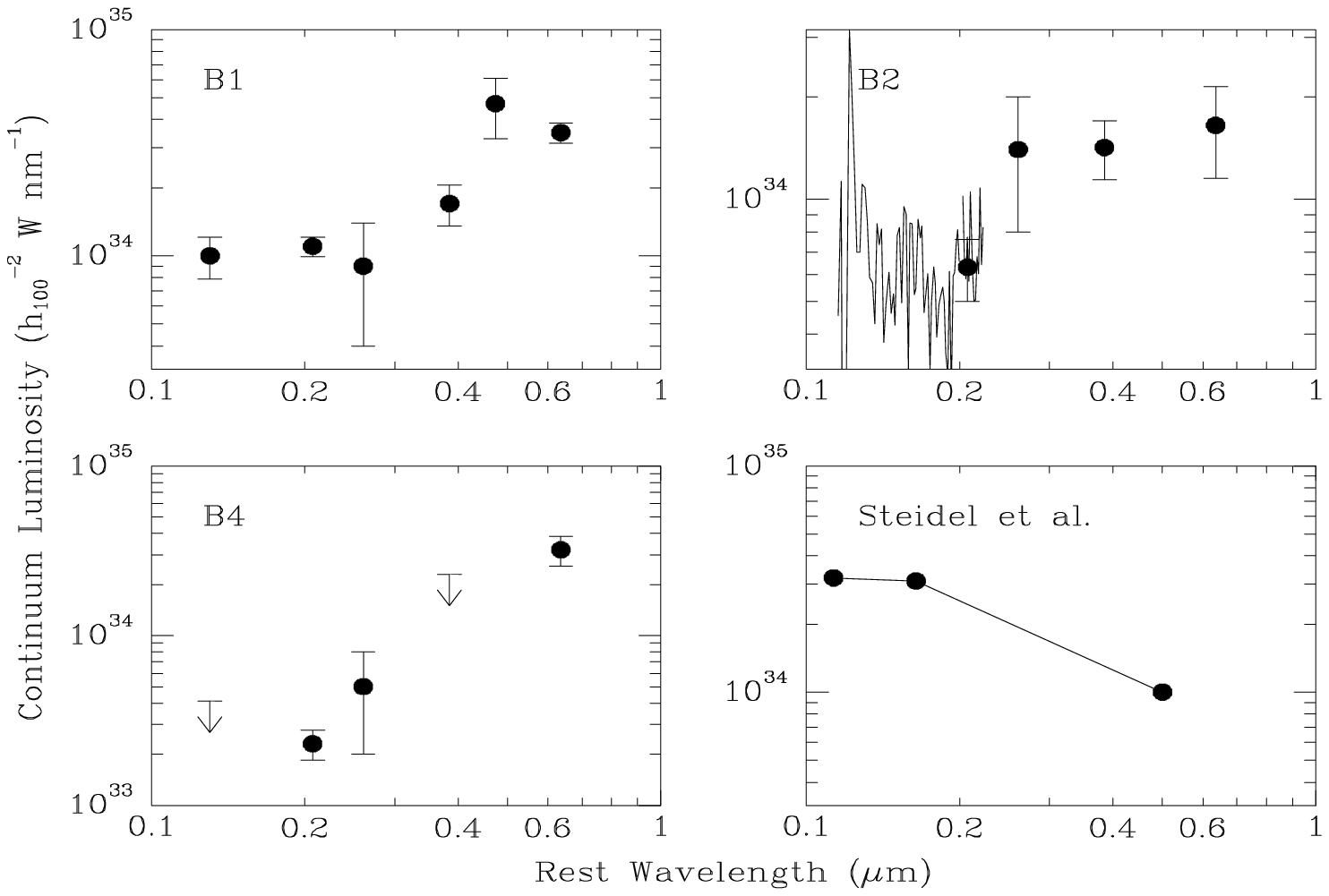}

\caption{Continuum spectral energy distributions of the
three Ly$\alpha$ galaxies, compared with a composite spectral energy
distribution for the galaxies in Steidel et al. (1996)
with listed $K$-band magnitudes. $1 {\rm W\ nm}^{-1} \equiv 10^{6} {\rm 
erg\ s}^{-1}{\rm \AA}^{-1}$\label{Fig2}}

\end{figure}

The properties of these three confirmed Ly$\alpha$ emitting galaxies at
$z=2.38$ are listed in Table~\ref{prop}, and their spectral energy
distributions are shown in Fig~\ref{Fig2}. The redshifts of all three
galaxies agree with each other, and with the redshift of FH93's\markcite{FH93} 
absorption-line cluster to within $\sim 600 {\rm km\ s}^{-1}$, and all lie
within one square arcmin ($1\arcsec \sim 5$kpc in proper coordinates).
The galaxies thus form a group of size $\sim 1$ Mpc; if they were
randomly distributed in the surveyed volume the probability of
finding them so close would only be $10^{-5}$.

Fig~\ref{Fig2} shows that all three Ly$\alpha$ galaxies have extremely
red colors, even after the removal of Ly$\alpha$ and H$\alpha$ line
emission from the $B$ and $K$ bands. B4 is spatially unresolved in all 
bands, but both B1 and B2 are marginally spatially resolved in the K-band;
we defer a discussion of their morphologies pending upcoming Hubble Space 
Telescope observations.

\section{Discussion}

\subsection{Emission-lines}

Our galaxies have much stronger UV line emission than most of those of
Steidel et al. (1996)\markcite{S96}. Several pieces of evidence suggest
that this strong line emission is powered by photoionization from
active nuclei, and not by young stars. The strength of the high ionization 
lines (\ion{C}{4} and \ion{He}{2}) is similar to that seen in 
radio galaxies, and implies that the emission-line
gas is exposed to a hard ionizing continuum. The Ly$\alpha$ equivalent 
widths are close to the maximum that can be explained by stellar ionization
(Charlot \& Fall\markcite{cf93} 1993). In B1 the Ly$\alpha$ emission
comes from a much more extended region than the continuum
emission (F96\markcite{F96}), and in B2, both the Ly$\alpha$ and
C~IV emission are displaced along the spectrograph slit away from the continuum 
source by $\sim 2\arcsec$ ($\sim 10$ kpc). This suggests that
the ionization source is physically separated from the emission-line
region, which is most easily explained if the former is an active nucleus.

These Ly$\alpha$ sources are not however conventional QSOs: their 
Ly$\alpha$ equivalent widths ($> 10$ nm rest-frame) are higher than those of 
typical QSOs ($\sim 5$ nm, 
Francis\markcite{F93} 1993), their velocity widths ($< 1500 {\rm km\ 
s}^{-1}$) are narrower than those of typical QSOs ($\sim 5000 {\rm km\ 
s}^{-1}$, Francis, Hooper \& Impey\markcite{fhi93} 1993), and their
continuum slopes are much redder ($B-K > 5$) than those of typical
QSOs ($B-K \sim 2.5$, Francis\markcite{f96} 1996). The Ly$\alpha$ fluxes
are however typical of the {\em extended\ } line emission around some 
high redshift quasars (eg. Heckman et al. 1991\markcite{hk91}) and radio 
galaxies (McCarthy 1993\markcite{mc93}). We therefore suggest that these three
galaxies contain QSO IIs; radio-quiet active nuclei, which are concealed 
from our 
line of sight, perhaps by a dusty torus, ie. high redshift analogues of 
Seyfert II galaxies. Ionizing radiation from the concealed active nucleus 
does not reach us, but ionizes part of the gaseous environment of the host
galaxy. If our sources have the same ratio of extended Ly$\alpha$ flux
to continuum flux found in Heckman et al.'s (1991)\markcite{hk91} sample of 
radio-loud quasars, the unobscured active nucleus would have an absolute 
$B$ magnitude of $\sim -24$; ie. typical of weak QSOs, rather than Seyfert 
galaxies. This should be regarded as a lower limit to the QSO luminosity,
as radio-quiet QSOs typically produce no detectable extended Ly$\alpha$ 
emission.
 
In both B1 and B2, the emission-line spectra (Fig~\ref{Fig1}) were 
obtained $> 1\arcsec$ from the continuum emission. For B1 the slit passed 
through the peak of the Ly$\alpha$ emission, which lies $\sim 1.5\arcsec$ to 
the SW of the continuum source (F96), and no continuum emission is seen in the
spectrum. For B2, the Ly$\alpha$ and C~IV emission were displaced $\sim
1\arcsec$ to the north of the top of the continuum emission, itself
extended and clearly seen in the spectrum. The observed velocity widths of
Ly$\alpha$ in both sources ($\sim 500 {\rm km\ s}^{-1}$) cannot,
therefore, be due to the motion of emission-line clouds in the
potential of a central black hole.
If the gas is in virial equilibrium with the host galaxy mass, we derive
host galaxy masses of $\sim 10^{12} M_{\sun}$. If the host galaxies
are not this massive, however, some form of $\sim 10$ kpc scale radio-quiet 
wind or jet, or motion of the gas in a massive cluster potential, is required 
to give the observed velocities. 

\subsection{Continuum Emission}

All three of our galaxies have much redder optical/near-IR continuum
colors than those of Steidel et al. (Fig~\ref{Fig2}). They are also
much redder than the Ly$\alpha$ emitting galaxy discovered by
Djorgovski et al. (1996)\markcite{dj96}; no near-IR photometry is
published for the other known Ly$\alpha$ selected high redshift galaxies
(eg. Lowenthal et al. 1991\markcite{LO91}, Pascarelle et al. 
1996\markcite{PA96}). The high redshift galaxies studied by
Malkan, Teplitz \& McLean (1996)\markcite{mal96} and Ellingson et al.
(1996)\markcite{ell96} also have bluer optical/near-IR colors. The
colors of our galaxies are, however, typical of many high redshift
radio galaxies (eg. Lilly 1989\markcite{lil89}), though due to our
narrow-band H$\alpha$ imaging, we can remove the (small) emission-line
contribution to our continuum colors, which is important in some radio 
galaxies (eg. Eales et al. 1993\markcite{eal93}). Our galaxies are not as 
red as those of Hu \& Ridgway (1994)\markcite{hr94}.

The red colors of our galaxies could be explained in three ways: by 
an old stellar population, dust obscuration, or reddened AGN light.
Previous studies demonstrate that it is very hard to discriminate
between these possibilities solely on the basis of broad-band photometry
(eg. Lilly 1989\markcite{lil89}, Chambers 
\& Charlot 1990\markcite{cc90}, Bithell \& Rees 1990\markcite{br90}, 
Graham \& Dey 1996\markcite{gd96}, Dunlop et al.
1996\markcite{du96}).

As both B1 and B2 are spatially extended in the K-band, any AGN contribution
to their red colors must be scattered. Polarimetry and HST imaging will
indicate if this is significant in any of our galaxies. If the observed
continuum emission is due to starlight, the galaxies could, 
at one extreme, be forming stars at $\sim 10^4 M_{\sun}{\rm yr}^{-1}$, with
substantial dust obscuration ($E(B-V) \sim 1$), or at the other
extreme, be dust free, massive  ($> 2 \times 10^{11} M_{\sun}$), and
old ($> 1$ Gyr). These limits were calculated using the spectral
synthesis models of Bruzual \& Charlot (1995, see Bruzual \&
Charlot 1993\markcite{BC93}), and dust extinction law of Calzetti
et al. (1994)\markcite{cal94}.

Producing red, spatially extended AGN light requires fine-tuning of the 
scattering geometry and viewing angle, which makes the presence of three 
such galaxies somewhat unlikely.
The strong Ly$\alpha$ flux also suggests that these galaxies are not very
dusty, as Ly$\alpha$ photons are very easily destroyed by even moderate
amounts of dust. We therefore tentatively suggest that the red colors of
our galaxies are produced by an old stellar population. This would, however,
require the formation of mass concentrations of $\gg 10^{12} M_{\sun}$
at redshifts above 5; hard to explain in most cosmological models
(eg. Kashlinsky \& Jiminez 1997\markcite{kj97}).

\section{A New Population of High Redshift Galaxies? \label{newpop}}

How common could red galaxies be at high redshifts? Their red colors
and clustering suggest that
they might be the ancestors of some low redshift elliptical galaxies.
If an appreciable fraction of ellipticals completed the bulk of their
star formation before $z \sim 5$, as suggested by observations at
intermediate redshifts (eg. Schade et al. 1996\markcite{sc96}), the
co-moving space density of red galaxies at $z \sim 2$ could be much higher 
than the observed upper limits on the co-moving densities of radio- and
Ly$\alpha$ emitting galaxies. The difference could be made up by a population 
of red galaxies without active nuclei, which would not be found in radio or 
emission-line surveys.

Are broad-band optical surveys such as that of Steidel et al. 
(1996)\markcite{S96} and the Hubble Deep Field capable of finding this
hypothetical population of red high redshift galaxies without active nuclei?
Consider a model in which half of all the high redshift galaxies with a 
given rest-frame $V$-band magnitude are blue (with the colors of the 
Steidel et al. galaxies) and half are red (with the colors of our galaxies). 
The ratio of observed-frame optical flux to rest-frame optical flux of the 
blue galaxies is an order of magnitude greater than that of the red galaxies. 
Thus to be included in a sample with a magnitude limit in the observed-frame 
optical, a red galaxy will
have to be an order of magnitude more luminous in the rest-frame $V$-band
than a blue galaxy. If the high redshift $V$-band luminosity function
is similar to a Schechter function (Schechter 1976\markcite{SH76}), the
space density of the red galaxies in a optically magnitude limited sample 
will thus be only $\sim 10^{-4}$ that of blue galaxies. As current optically
selected surveys have sizes of only $\sim 10^2$ galaxies, they would not 
therefore contain any red galaxies. Some recent IR selected galaxy surveys
are, however, finding sources that could be red, high redshift galaxies
(Moustakas et al. 1997\markcite{mou97}).
 
In conclusion, we have detected a group of three radio-quiet high redshift
galaxies with strong emission-lines and red continuum colors. The 
emission-lines are probably caused by photoionization by a concealed 
radio-quiet active nucleus, while the red colors could be due to some 
combination of dust, scattered AGN light and an old stellar population.
Equally red galaxies without active nuclei could be an important
population in the high redshift universe.

We wish to thank Bob Dean, Karl Glazebrook, and all the staff at the AAT, 
for their able assistance, and Alistair Walker for obtaining the CTIO service
imaging.

\begin{deluxetable}{lccc}
\tablecolumns{4}
\tablecaption{Galaxy Properties \label{prop}}
\tablehead{
\colhead{}   &
\colhead{B1} &
\colhead{B2} &
\colhead{B4} }
\startdata
RA (B1950)\tablenotemark{a}   & 21:39:16.38 & 21:39:18.56 & 21:39:20.98 \\
DEC (B1950)\tablenotemark{a}  & $-$44:34:12.2 & $-$44:34:45.0 & $-$44:34:00.8 \\
Ly$\alpha$ Flux &  & & \\
($\times 10^{-19} {\rm W m}^{-2}$) & $8.0 \pm 0.5$ &
$2.1 \pm 0.3$ & $1.8 \pm 0.3$ \\
Ly$\alpha$ Rest-frame & & & \\
Equivalent Width (nm) & $22 \pm 4$ & $> 10$ & 
$> 12$ \\
H$\alpha$ Flux &  & & \\
($\times 10^{-19} {\rm W m}^{-2}$) & $9.2 \pm 1.5$ &
$< 6.6$ & $4.0 \pm 1.4$ \\
Ly$\alpha$/\ion{C}{4} & 7 & 8 & 4.4 \\
Ly$\alpha$/\ion{He}{2} & \nodata & $> 16$ & 3.3 \\
Ly$\alpha$ velocity & & & \\
width (FWHM, ${\rm km s}^{-1}$) & 600\tablenotemark{c} &
450\tablenotemark{c} & 1200 \\
\cutinhead{Line-subtracted continuum colors}
$B$ & $25.9 \pm 0.2$ & $> 26.6$ & $>26.9$ \\
$R$ & $24.2 \pm 0.1$ & $24.8 \pm 0.2$ & $25.9 \pm 0.5$ \\
$I$ & $23.8 \pm 0.5$ & $23.3 \pm 0.5$ & $24.3 \pm 0.5$ \\
$J$ & $21.8 \pm 0.2$ & $22.0 \pm 0.2$ & $> 21.5$ \\
$H$ & $19.9 \pm 0.3$ & \nodata & \nodata \\
$K$ & $19.0 \pm 0.2$ & $19.9 \pm 0.3$ & $19.1 \pm 0.2$ \\
\enddata
\tablenotetext{a}{Position is for the centroid of the Ly$\alpha$ flux.}
\tablenotetext{b}{Upper and Lower limits are $3 \sigma$.}
\tablenotetext{c}{Measured from the higher resolution spectra of F96.}
\end{deluxetable}

\end{document}